\documentclass{article}
\usepackage[utf8]{inputenc}
\usepackage{graphicx}
\usepackage{amssymb}
\usepackage{epstopdf}
\usepackage{authblk}

\title{Estimating the impact of preventive quarantine with reverse epidemiology}
\author[1]{Jacopo Grilli}
\author[1]{Matteo Marsili}
\author[2]{Guido Sanguinetti}
\affil[1]{\footnotesize The Abdus Salam International Center for Theoretical Physics, Trieste, Italy}
\affil[2]{\footnotesize International School for Advanced Studies (SISSA), Trieste, Italy}
\date{April 2020}

\begin{document}
\maketitle
\begin{abstract}
The impact of mitigation or control measures on an epidemics can be estimated by fitting the parameters of a compartmental model to empirical data, and running the model forward with modified parameters that account for a specific measure. This approach has several drawbacks, stemming from biases or lack of availability of data and instability of parameter estimates. Here we take the opposite approach -- that we call reverse epidemiology. Given the data, we reconstruct backward in time an ensemble of networks of contacts, and we assess the impact of measures on that specific realization of the contagion process. This approach is robust because it only depends on parameters that describe the evolution of the disease within one individual (e.g. latency time) and not on parameters that describe the spread of the epidemics in a population. 
Using this method, we assess the impact of preventive quarantine on the ongoing outbreak of Covid-19 in Italy. This gives an estimate of how many infected could have been avoided 
had preventive quarantine been enforced at a given time.
\end{abstract}

\section{Introduction}

The spread of a pandemic such as Covid-19 may not be contained efficiently by social distancing measures alone \cite{Wang}. At least in part, this is because social distancing may not be possible in the household or in other circumstances (e.g. in a hospital). In addition, lock-down of entire countries is clearly unsustainable in the long run, for several reasons. 
Social distancing reduces the number of contacts of individuals, unconditionally on their morbidity state. Preventive quarantine is an additional measure where quarantine is applied to all the contacts of an infected individual as soon as the latter is tested positive. This may not stop contacts of the infected individual from developing the infection, but it closes the possibility that, if the contact develops the infection, he or she will go on to infect further people. 

This measure -- also called Centralised Quarantine -- has been enforced in China \cite{Wang} by establishing centralised quarantine centres, also\footnote{Contacts of positive cases and symptomatic cases were quarantined in different facilities \cite{Wang}.} for close contacts of confirmed cases, to prevent further spread of the epidemics in case they also turned Covid19 positive. 

Ref. \cite{Wang} took a model based approach to study the impact of these measures {\em a posteriori} in the Wuhan epidemics. During the epidemics, this approach is not feasible, because data is scarce and unreliable, and consequently estimates of model parameters are affected by considerable uncertainty. We apply a model-free strategy for using data available up to now (or possibly further extrapolated with plausible scenarios) to understand how the dynamics of the epidemics would have changed if preventive quarantine had been enforced at a particular time in the past. The idea is that the dynamics of the epidemics can be reversed backward, thereby constructing a probabilistic network of contagion contacts in the population of confirmed cases. For each such reconstruction, the epidemics can be run forward again, with preventive quarantine enforced. This provides a picture of how much preventive quarantine would have affected the epidemics, had it been implemented at a given time. 

\section{Reversed epidemics}

For preventive quarantine to work, it is essential to be able to trace contacts. Let us assume that the fraction of infected people who know by whom they have been infected is $p$. For countries where contact tracing apps have been introduced, $p$ can be taken as a measure of the fraction of individuals that have that app installed and active on their mobile phone. Alternatively, contact tracing can be done by interviewing positives; a recent online survey of Spanish respondents, suggests that $p$ be as high as 80\% on the basis of individual interviews \cite{NuriaOliver}. Ferretti {\em et al.} \cite{Ferrettieabb6936} estimate that if the adoption of a contact tracing app covers more that 70\% of a population, then an epidemic with the traits of SARS-CoV-2 can be tamed.

We assume that, if A knows that B infected him/her\footnote{By this we mean that A knows the identity of B. This does not count cases as, for example, when A knows that s/he has been infected by a client of the grocery shop he/she visited. It is unclear whether the figure of 80\% reported in \cite{NuriaOliver} includes such cases or not}, then B must have been tested positive in a previous day. We're also going to assume that, in this case, B also knows that he was in contact with A. With contact tracing apps, these are reasonable assumptions, otherwise this may be reasonable if the relationship between A and B is a significant one.

Let $\mathcal{I}_t$ the set of new positives at time $t$. Let $\mathcal{P}_t\subseteq\mathcal{I}_t$ be the subset of infected at time $t$ that know by whom they were infected. For each individual $i\in \mathcal{P}_t$ we can simulate her previous history of contagion: she got infective (with or without symptoms) at time $t-T_{i}$ and the contact that passed them the infection occurred at time $t-T_i-L_i$, where $T_i,L_i=1,2,\ldots$ are times of infection and latency and can be drawn from a  distribution\footnote{Uppercase letters are used for random variables.}. There is ample converging evidence on the expected values of $T_i$ and $L_i$ for Covid-19 (see \cite{pellis2020challenges} for a review), and also their distribution has been estimated from data \cite{Linton}. We assume that all those tested positive at time $t$ do not infect others at time $t$ or at later times, i.e. they are immediately quarantined (or hospitalised). This may not be true in reality.

From a time series of positive new cases, we can reconstruct at each time $\tau\le t$ a population $\mathcal{C}_\tau$ of contacts, which is the set of people that got infected at time $\tau$, and the population of $\mathcal{E}_\tau$ of infective, but not yet detected, individuals. $\mathcal{E}_\tau$ contains all those individuals for which $t_i-T_i\le\tau<t_i$, where $t_i$ is the time where $i$ was tested positive. For each $i\in \mathcal{C}_\tau$ we can draw at random a link (i.e. a contagion event) to a $j\in\mathcal{E}_\tau$ and reconstruct the contact network at time $\tau$. This construction\footnote{This corresponds to sampling the maximum entropy ensemble of contact networks consistent with the data. This is appropriate in the absence of further information on the statistics of contacts. The approach can be refined to include additional information on contact distributions.} gives a possible history of how the contagion has spread up to time $t$. For all $i\in\mathcal{E}_\tau$ we define $\partial_{i,\tau}$ as the set of all $j\in\mathcal{C}_\tau$ that were infected by $i$ at time $\tau$. It is important to remark that this is only a plausible history of the contagion; nevertheless, if the statistics of the latency/ detection times and traceability are reasonable, we expect that conclusions drawn from multiple simulated histories will provide a faithful representation of the average behaviour of the epidemic.

Now let us assume that at time $t_{pq}$ a preventive quarantine measure is put in place. This implies that all individuals that got in contacts of all $i\in\mathcal{I}_\tau$ since $\tau-t_Q$ are isolated (with $t_Q=14$ days). We generate a set $\mathcal{Q}_\tau$ that contains all individuals in quarantine
\begin{equation}
\label{ }
\mathcal{Q}_\tau=\mathcal{Q}_{\tau-1}\bigcup\left[\bigcup_{i\in\mathcal{P}_\tau}\bigcup_{\tau'=\tau-t_Q}^\tau\partial_{i,\tau'}\right].
\end{equation}
In words, $\mathcal{Q}_\tau$ contains all individuals in quarantine the day before, plus all those that where in contact with all $i\in \mathcal{P}_\tau$ in the previous $t_Q$. Notice there is no need to remove people from $\mathcal{Q}_\tau$, since we're considering only positives that stay in quarantine until they become negative.

As a consequence all contagion events involving $i\in\mathcal{Q}_\tau$ are eliminated. After this is done, the network of contagion is reconstructed on all the remaining contagion events, a new population $\mathcal{P}^Q_\tau$ is reconstructed.

\section{Results}

\begin{figure}[ht]
\centering
\includegraphics[width=0.48\textwidth,angle=0]{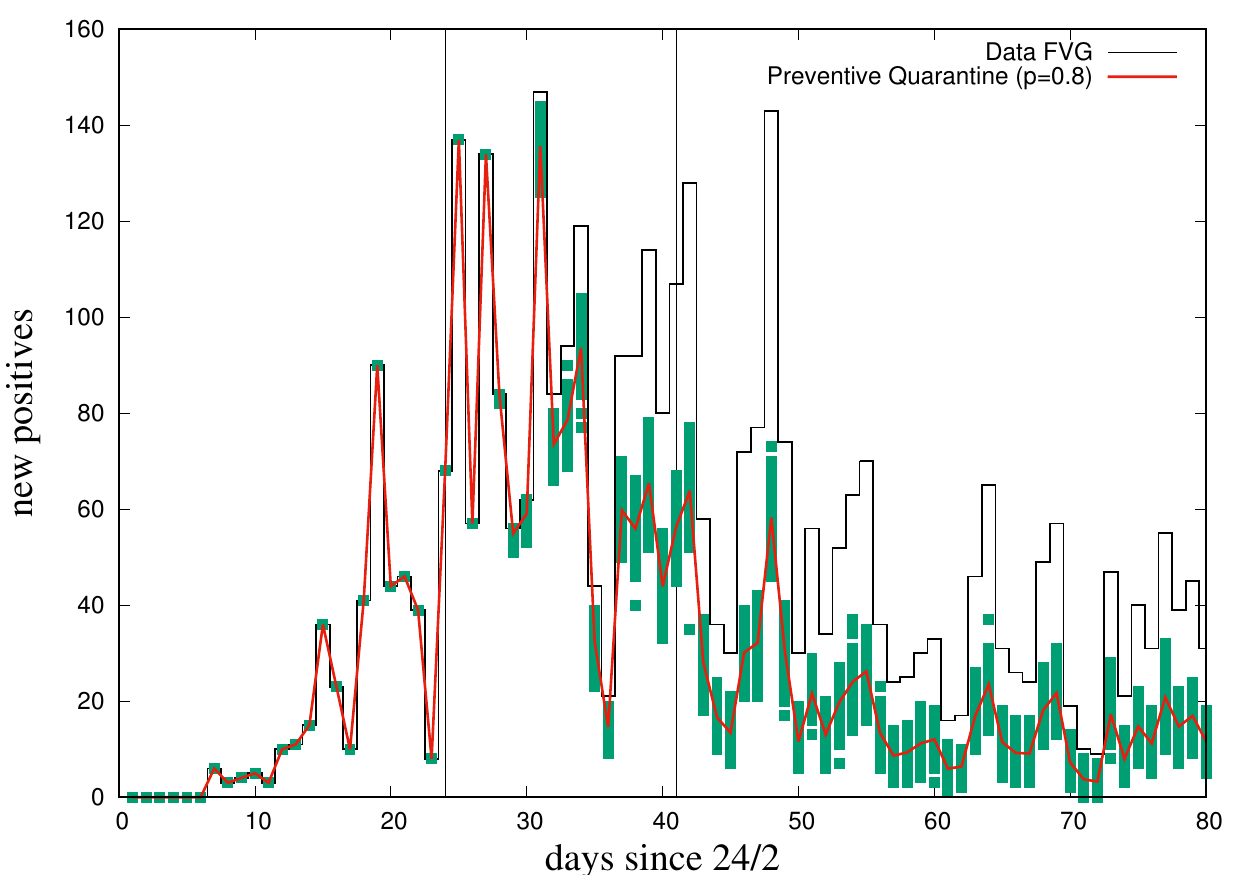}
\includegraphics[width=0.48\textwidth,angle=0]{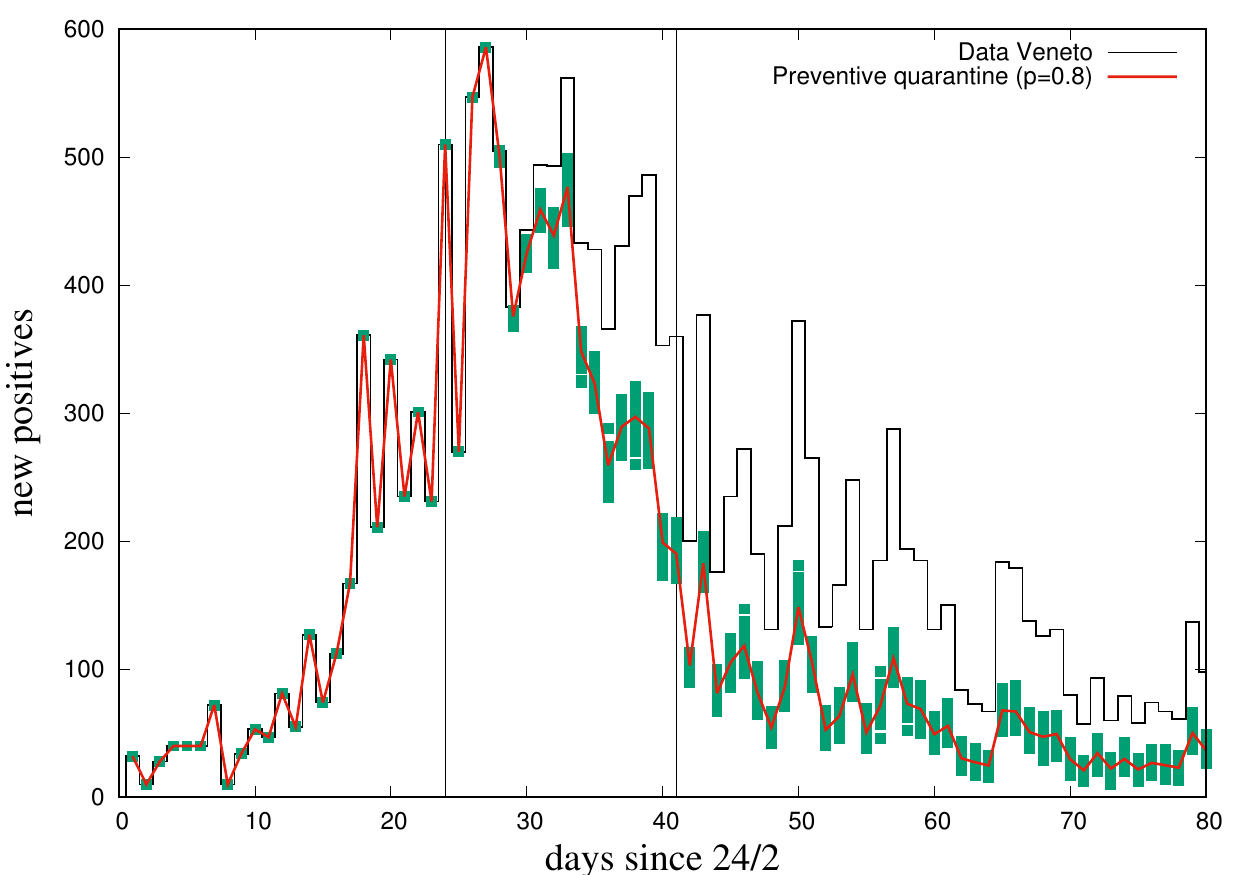}
\includegraphics[width=0.48\textwidth,angle=0]{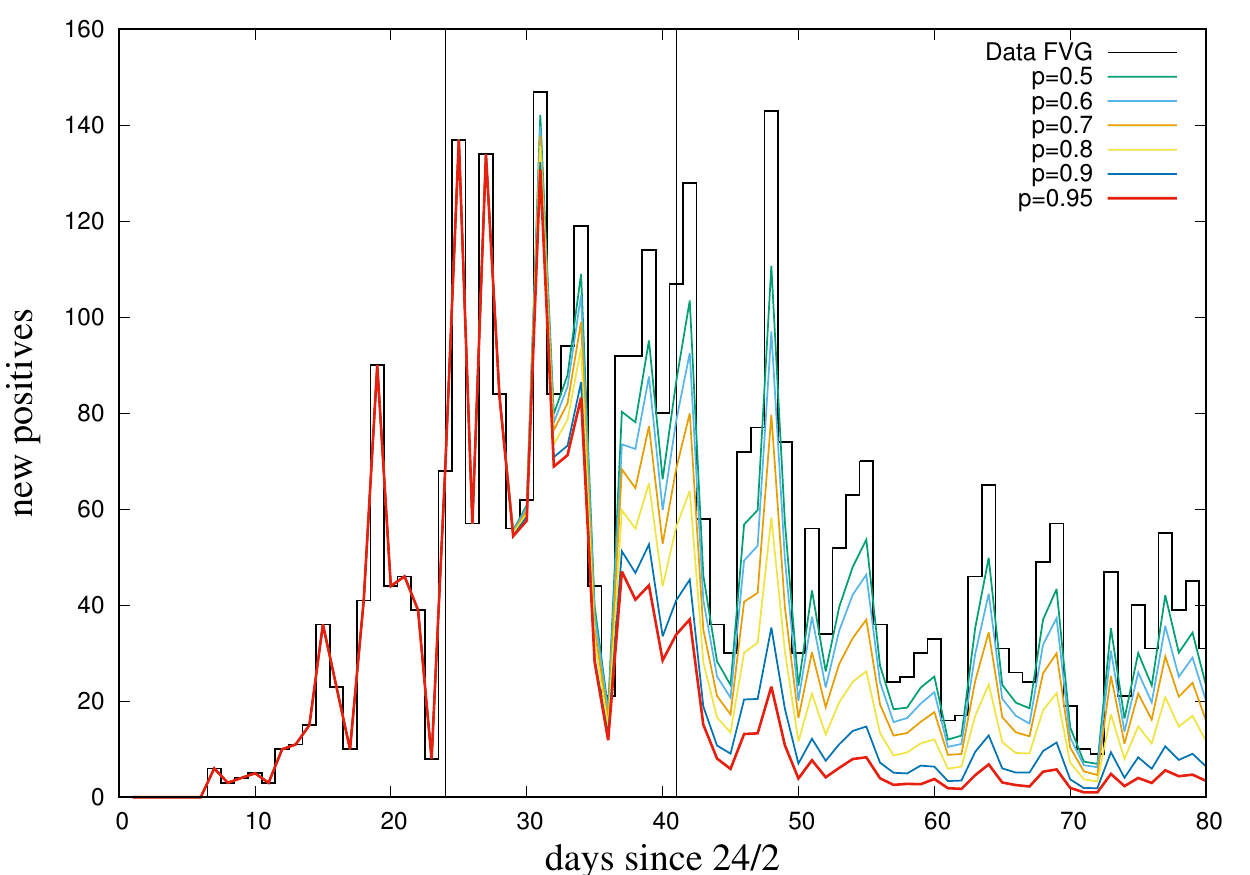}
\includegraphics[width=0.48\textwidth,angle=0]{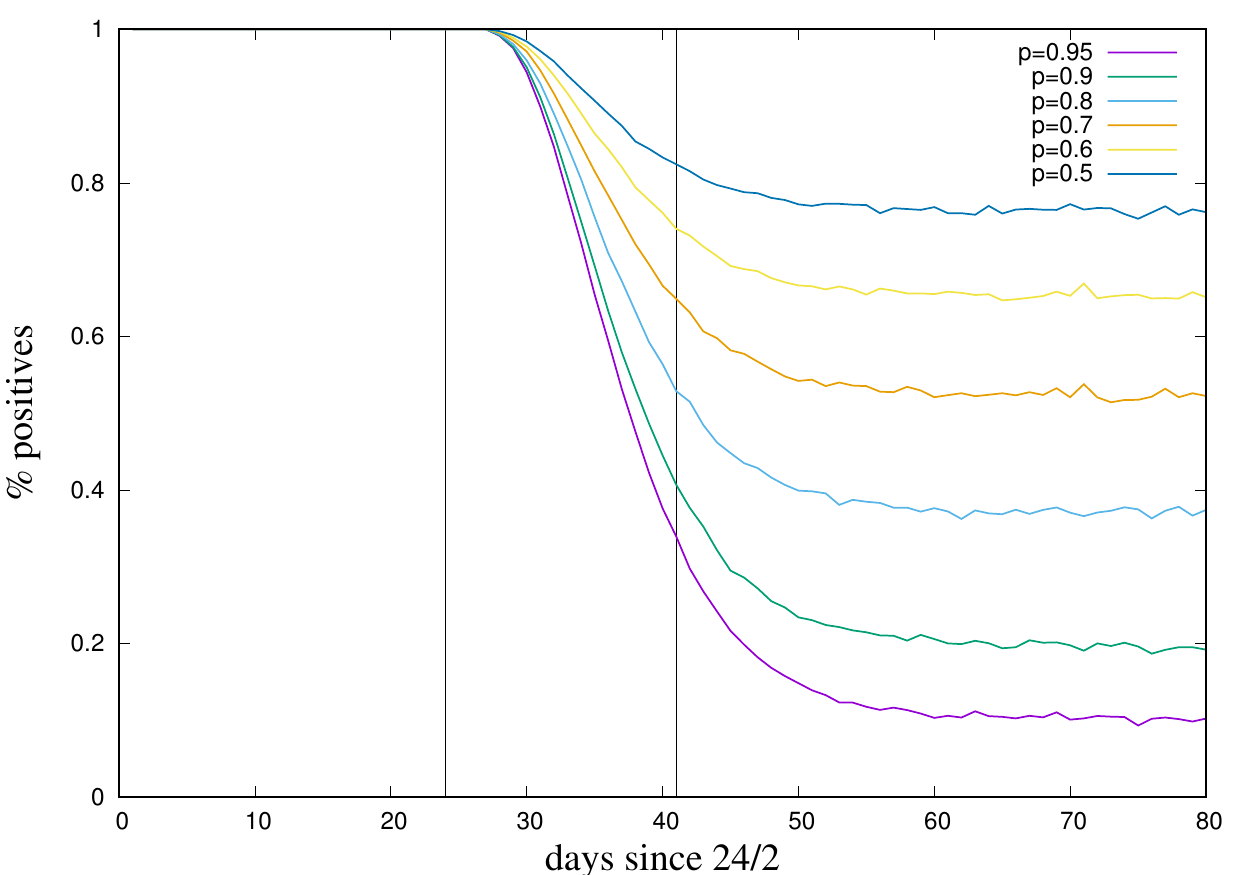}
\caption{\label{Fig1} Top: New positive Covid-19 cases (black lines) for the FVG (top left) and Veneto (top  right) regions. The data up to April $4^{\rm th}$ (vertical line at day 41) is continued by simulating a possible scenario. This dataset is compared with the new positives under a preventive quarantine measure ($p=0.8$) enforced on March $18^{\rm th}$ (vertical line at day 24) in 100 independent runs of the reconstructed epidemics (green dots). Averages are reported as red lines. Bottom left: Average new positives in 100 runs of he reconstructed epidemics with preventive quarantine for different values of $p$ for the FVG Region. Bottom right: Fraction of the new positives that should be dealt with with preventive quarantine for different values of $p$ for the Veneto Region.}
\end{figure}

The result of the reconstruction\footnote{The code used for the simulations is available under request at {\tt marsili@ictp.it}.} of the epidemics is shown in Figure \ref{Fig1}. Data from Friuli Venezia Giulia (FVG) and Veneto regions are compared to what would have happened on average if preventive quarantine had been enforced on March $18^{\rm th}$ (day 24). We extend the data up to April $5^{\rm th}$ with a possible scenario of how the epidemics may evolve under a slowly varying downward trend \footnote{Notice that these are not predictions; they are simply illustrative of a possible forward history of the epidemic and the relative effect that preventative quarantine would have.}. This sheds light on the effects of preventive quarantine on a longer timescale. We assumed a geometric distribution for the distributions of $T_i$ and $L_i$, with $E[T_i]=3$ and $E[L_i]=5$.

After sufficient time, the ratio between the predicted new positives under preventive quarantine and the observed ones approach a constant. The value of this proportion can be easily calculated by considering that new cases under preventive quarantine are possible only if a person does not know by whom was infected (event that happens with probability $p$) or if she knows that was infected by someone who was not under preventive quarantine (which happens with probability $p(1-p)$). For large times the ratio of new positive with and without preventive quarantine tends to $1-p^2$.

\section{Discussion}

There are a number of aspects worth mentioning:
\begin{itemize}
  \item Individuals $i\in\mathcal{I}_\tau/\mathcal{P}_\tau$ who don't know whom they've been infected by are not removed. This is a clear incentive for adopting apps at individual level. We assume that individuals who don't know whom they are infected by, are infected by individuals that do not belong to the population of detected positives.
  \item We assume that if $i$ knows that he/she was infected by $j$ then $j$ knows, at the time of being tested positive, that he/she had a contact with $i$. This may not be true based on questionnaire, but it is reasonable if apps for contact tracing are adopted.
  \item The set $\mathcal{Q}_\tau$ only includes positives in quarantine. It does not include those who got in contact with a positive case, but who were not infected. So the set of quarantined people should be much larger. Ten times the number of positives may be a a conservative measure.
  \item The reverse epidemics approach relies on parameters of the infection dynamics (the distributions of $T_i$ and $L_i$) that have been estimated empirically \cite{Wang,pellis2020challenges}. In particular, the approach is independent of the infection rate, percentage of asymptomatic cases and other parameters that make predictions on the basis of forward epidemiological models very unstable. 
  \item Effects of preventive quarantine are weak in the first week (see Fig. \ref{Fig1}) due to the lags related to incubation period and detection. A significant effect is apparent by the end of the second week. The full effect sets in after the third week.
  \item The effects of preventive quarantine depend strongly on $p$, the efficiency with which contacts are monitored. An efficiency $p=0.8$ in tracing contacts, that according to a recent study should be achievable by interviewing positive cases \cite{NuriaOliver}, can reduce the size of the epidemics by 60\%. An efficiency of 95\%, such as that reported by Wang {\em at al.} \cite{Wang} for Wuhan, can reduce almost to 10\% of what they would have been. If mobile apps will be adopted the level of compliance of users will be critical to ensure that preventive quarantine is effective. 
  \item Our analysis permits us to estimate the effects of preventive quarantine on the tested positive population in a counterfactual scenario. In reality, preventive quarantine would also isolate many infected individuals who {\it would not} go on to develop symptoms (and hence would be unlikely to be tested) but who would contribute to further spread the disease.
  The isolation of non-tested individuals would contribute, in the most optimistic scenario, to a reduction of basic reproductive number of a factor $1-p^2$, if all the contacts between people that know who they were infected by are eliminated. This optimistic reduction of the basic reproductive number would correspond to an exponential decrease of the ratio between new cases with and without preventive quarantine.  
  The protective potential of preventive quarantine might be therefore between this latter optimistic scenario and the more pessimistic one that we discussed in this paper and illustrated in Figure 4.
\end{itemize}

The reverse epidemiology approach bears similarity with the reconstruction of phylogeny from generic data (see e.g. \cite{phylogeny}) and the counterfactual analysis of disruption events in other complex systems (see e.g. \cite{silva2015predicting}). We are not aware of any such approach to the study of epidemic outbreaks, which we believe is a very promising one.

\bibliographystyle{acm}
\bibliography{bibcovid}

\end{document}